\documentstyle[11pt,aaspp4]{article}

\def\iras{{\it IRAS\ }}
\def\mpc{${h^{-1}\rm Mpc}$}
\begin{document}
\title{The Signature of a Correlation between  
$>10^{19}{\rm eV}$ Cosmic Ray Sources and Large Scale Structure}
\author{Eli Waxman and  Karl B. Fisher}
\affil{Institute for Advanced Study, Princeton, NJ 08540}
\author{Tsvi Piran}
\affil{Racah Institute for Physics, The Hebrew University, Jerusalem,
91904, Israel}
\authoremail{waxman@sns.ias.edu}

\begin{abstract}

We analyze the anisotropy signature expected 
if the high energy (above $10^{19}$eV) cosmic ray (CR) 
sources are extra-Galactic and trace the distribution of luminous matter on 
large scales.  We investigate the dependence of the anisotropy 
on both the relative bias between the CR sources and the galaxy distribution 
and on the (unknown) intrinsic CR source density. 
We find that the expected anisotropy associated 
with the large scale structure (LSS) 
should be detected once the number of CR events observed above 
$10^{19}{\rm eV}$ is increased by a factor of $\sim10$. This would require
$\sim30$ observation-years with existing experiments, but less then 
$1$ year with the proposed $\sim5000\ {\rm km}^2$ Auger detectors. 
We find that the recently reported concentration of the Haverah Park 
CR events towards the super-galactic plane is not consistent with the known
LSS. If real, the Haverah Park result suggests that the CR sources are much 
more concentrated towards the super-galactic plane than the known LSS.
Our results are not sensitive to the number density of CR sources. 
We show that once the number of detected events is increased by a factor 
of $\sim10$, the number density would be strongly constrained by
considering the probability for having repeating sources. 

\end{abstract}

\section{Introduction}

Recent cosmic ray observations, reported by the Fly's Eye (\cite{Fly}) 
and by the AGASA (\cite{AGASA}) experiments, show two major features in the 
cosmic ray (CR) energy spectrum above $10^{17}{\rm eV}$. First, 
a break in the shape of the spectrum is observed at 
$\sim5\times10^{18}{\rm eV}$. Second, the CR composition changes 
from being predominantly heavy nuclei below the break to 
light nuclei above the break. Coupled with the lack of anisotropy, 
that would be expected if the CRs above $10^{19}{\rm eV}$ were protons
produced in the Galaxy, these features strongly suggest that the CR flux
above $10^{19}{\rm eV}$ is dominated by an extra-Galactic component of 
protons. This view is supported by the fact that the CR spectrum above 
$\sim2\times10^{19}{\rm eV}$ is consistent with a cosmological distribution 
of sources, with a power law generation spectrum ${\rm d}\ln N/{\rm d}
\ln E\simeq-2$ (\cite{Wb},c) 
(below $\sim2\times10^{19}{\rm eV}$ a significant
contribution from iron cosmic rays from Galactic sources is likely to
be present; Bird {\it et al.} 1994, Waxman 1995b).

If the particles observed are indeed protons of extra-Galactic origin, and
if their sources are correlated with luminous matter, then the 
inhomogeneity of the large scale galaxy distribution, on 
scales $\lesssim100$\mpc, should be imprinted on the CR arrival directions. 
In this paper, we examine the expected anisotropy signature
if the CR sources trace the large scale structure (LSS),
and investigate its dependence on the relative bias 
between the CR sources and the galaxy 
distribution and on the (unknown) intrinsic CR source density. The galaxy  
distribution is derived from the \iras 1.2 Jy redshift
survey (Fisher {\it et al.} 1995). We find that the expected 
anisotropy associated with the LSS should be detected once the number of CR
events detected above $10^{19}{\rm eV}$ is increased by a factor of $\sim10$.
\cite{Stanev} have recently noted that
the arrival directions of $E>4\times10^{19}{\rm eV}$ CR events detected
by the Haverah Park experiment show a concentration in the
direction of the Supergalactic Plane (SGP) which is 
inconsistent with the hypothesis that the CR sources are 
distributed isotropically.  We confirm this result, but also 
find that the Haverah Park
CR distribution is unlikely to be explained by the hypothesis
that the CR sources trace the known LSS.

\section{The angular distribution of CRs in a cosmological scenario}

We consider a model where the CR flux is produced by a cosmological
distribution of proton sources that trace the large scale galaxy
distribution. We model the number density
of CR sources in a comoving volume element ${\rm d}V$ at redshift $z$
as drawn from a Poisson distribution with mean $\bar s(z)\,b[\delta\rho]\, 
{\rm d}V$, where $\bar s(z)$ is the average comoving number 
density of CR sources at redshift $z$ 
and $b$ is some (bias) functional of the local galaxy 
overdensity, $\delta\rho$. With this assumption, we are able to derive
in \S 2.1 a closed analytic expression for the probability
of observing a source which produces a specified number of events
at a detector for a given exposure and sky coverage.
In \S~2.2, we apply this formalism to generate
Monte-Carlo realizations of the CR distribution arising from
a population of sources which trace the LSS seen in the \iras
redshift survey.  
In \S~2.3, we analyze the anisotropy of CR arrival distribution and
discuss the implications of our results
for the observed anisotropy reported by \cite{Stanev}. We conclude in \S~3.

\subsection{Formalism}

As they propagate, high energy protons lose energy due to the
cosmological redshift and due to the production of pions and $e^+e^-$ pairs 
caused by interactions with microwave background photons. 
We denote by $E_0(E,z)$ 
the energy at which a proton must be produced at an epoch $z$ 
in order to be observed at present ($z=0$) with energy $E$. With this
definition, the CR flux above energy $E$, produced
by a source at redshift $z$, is given by 
$(1+z)\dot N[E_0(E,z)]/4\pi d_L(z)^2$, where $\dot N[E]$ is the rate 
at which the source produces protons above energy $E$ and $d_L(z)$ is the
luminosity distance. For simplicity, we assume that the sources are
identical, so that $\dot N$ is given by the present ($z=0$) source
number density, $\bar s_0$, and CR production rate per unit volume,
$\dot n_0$, as $\dot N[E]=\dot n_0(E)/\bar s_0$ 
(Note, however, that it is straightforward to include a source luminosity
function in our formalism). The number of detected CR events produced by
a source at redshift $z$ is modeled as a Poisson distribution
with mean 
\begin{equation}
\bar N(E,z)=\dot N[E_0(E,z)]{(1+z)AT\over4\pi d_L(z)^2}
={\dot n_0[E_0(E,z)]\over\bar s_0}{(1+z)AT\over4\pi d_L(z)^2},
\label{Nbar}
\end{equation}
where $A$ and $T$ are the detector area and observation time
respectively. 

If many faint sources produce the observed CR events, then a 
natural assumption would be that each source has a negligible
probability of producing more than one event. However, the
density of CR sources is as yet unknown and there is the
possibility that the observed events are comprised in part
of sources producing multiple events (repeaters). In the
following derivation, we present a formalism which enables 
the probability that a source will produce one or multiple events to be 
calculated for a specified source density; the familiar notion of
each source producing only one event is recovered in the limit of
infinite source density.

The number of CR {\it sources} in a comoving volume 
element ${\rm d}V$ at redshift 
$z$ is taken to be Poisson distributed with mean
$\bar S=\bar s(z)b[\delta\rho]{\rm d}V$. The probability that a source
would produce $i$ detected CR events above
energy $E$ is $P_i(E,z)=\bar N(E,z)^i\exp[-\bar N(E,z)]/i!$.
The probability that the number of sources producing $1$ events
is $n_1$,
the number of those producing $2$ events is $n_2$, and so on, is
\begin{eqnarray}
P(\{n_i\})&&=\sum_{S=N}^\infty {1\over S!}\bar S^S e^{-\bar S}
{S!\over(S-N)!\prod_i n_i!}\prod_i P_i^{n_i}\left(1-
\sum_i P_i\right)^{S-N}
\nonumber\\
&&=\left(\prod_i{1\over n_i!}\right)e^{-\bar S}\prod_i(\bar S P_i)^{n_i}
\sum_{S=N}^\infty {1\over(S-N)!}\left[\bar S\left(1-
\sum_i P_i\right)\right]^{S-N}
\nonumber\\
&&=\prod_i{1\over n_i!}(\bar S P_i)^{n_i}e^{-\bar SP_i}\quad ,
\label{prob}
\end{eqnarray}
where $N=\sum_i n_i$.
Thus, the number of sources producing $i$ events is Poisson distributed
with average $\bar SP_i$, and is statistically independent of the number
of sources producing $j\not=i$ events. This implies that the number $S_i$ of
sources producing $i$ events in a given angular region of the sky is Poisson
distributed with average given by the sum of $\bar SP_i$ over all relevant
volume elements, and that $S_i$ is statistically independent of $S_j$
for $i\not=j$. The average number per steradian
of sources producing $i$ events in the direction $\hat{\Omega}$
is given by
\begin{equation}
\bar S_i(E,\hat{\Omega})=\int {\rm d}z\, 
c\left|{{ {\rm d}t}\over{ {\rm d}z}}\right|\,
{d_L(z)^2\over1+z}\,\bar s(z)\, b[\delta\rho(z,\hat{\Omega})]\, P_i(E,z) 
\quad .
\label{Sbar}
\end{equation}

Equation~\ref{Sbar} makes it easy to compute the
mean number of sources producing a given number of events above a given
energy and forms the 
basis for the Monte-Carlo method described in the next section.
It is straight forward to show that 
the number per steradian of sources producing more than
one detected CR above energy $E$, i.e. the number of repeaters, is Poisson
distributed with average given by (\ref{Sbar}) with $P_i(E,z)=1-[1+
\bar N(E,z)]\exp[-\bar N(E,z)]$. 
The mean number per steradian of observed CR events is:
\begin{equation}
\bar N_o (E,\hat{\Omega})= 
{1\over4\pi}AT\dot n_0(E)
\int{\rm d}z\,c\left|{{\rm d}t\over{\rm d}z}\right|
{\bar s(z)\over\bar s_0}\,b[\delta\rho(z,\hat{\Omega})]\,
{\dot n_0[E_0(E,z)]\over\dot n_0(E)}\quad.
\label{S1bar}
\end{equation}

\subsection{Monte-Carlo analysis of CR arrival directions}

We now wish to exploit the formalism of the previous section
by making specific Monte-Carlo realizations of CR arrival
directions for a particular detector and exposure time.
Equation~\ref{Sbar} can be used to determine the mean
number of sources along a particular line of sight which produce a
given number of CR events, once the underlying cosmology, 
the source evolution $\bar s(z)$, the density field $\delta\rho$, and the 
biasing function, $b[\delta\rho]$, have been specified.
For the numerical calculations of the rest of the paper we adopt the following
cosmological scenario: flat universe with zero cosmological constant
($\Omega=1,\ \Lambda=0$), $H_0=100{\rm km\ sec}^{-1}{\rm Mpc}^{-1}$, and
non-evolving sources ($\bar s(z)=\bar s_0$). Our results are not sensitive
to the cosmological parameters ($\Omega,\ \Lambda$) and to source evolution,
since CRs of energy $E>4\times10^{19}{\rm eV}$ are produced by sources
at distances smaller than $500\ {\rm Mpc}$ (see Fig. \ref{fig1}), 
which are short on a 
cosmological scale. The function $E_0(E,z)$ is numerically
calculated under the above assumptions in a method similar to that described in
\cite{Wb}. The proton spectrum generated by the sources is assumed 
to be a power law, $\dot N(E)\propto E^{-1}$, which is 
consistent with the observed CR spectrum above $2\times10^{19}{\rm eV}$
(\cite{Wb}). 

Figure \ref{fig1} presents the fraction of the differential CR flux, 
that is contributed by sources at distances $<d$,
for a homogeneous distribution of CR sources,
for several values of $d$ (the results are obtained by numerically
integrating equation (\ref{S1bar}) with $b\equiv1$). The rapid energy
loss due to pion production gives rise to the sharp cutoff with
distance.  CRs with energies 
$\gtrsim8\times10^{19}{\rm eV}$ must originate 
from sources within $\lesssim100\ {\rm Mpc}$. High energy CRs
therefore offer a probe of the relatively local universe
which is uncontaminated by distant sources. 
Searching for the signal of anisotropy associated with 
the LSS involves a compromise between
diluting of the signal by working with lower energy events
and increasing the statistical uncertainty by restricting
the analysis to the rare high energy events.

We estimate the large scale galaxy density field
by smoothing the galaxy distribution of the 1.2 Jy \iras
redshift survey (Fisher {\it et al.} 1995) with a variable 
Gaussian filter with a dispersion given by the larger of 
$6.4$\mpc\ and the mean galaxy separation
(cf. Fisher {\it et al.} 1995). Variable smoothing is essential for 
maintaining high resolution nearby ($r\lesssim 25$ \mpc) while
simultaneously suppressing shot noise at
large distances ($r\gtrsim 75$ \mpc). The relatively
large local smoothing length of 6.4 \mpc\
(corresponding to the same volume as a top-hat filter of
radius 10 \mpc) was chosen so that the
smoothing length remains constant out to $\sim$ 100\mpc.
For distances $>200$ \mpc\ we have assumed a homogeneous 
($\delta\rho=0$) galaxy distribution. In the range $100<r<200$ \mpc,
the smoothing length increases from 15 \mpc\ to about 45 \mpc.
At first glance, this degradation in resolution might appear
worrisome. We tested the robustness of our results by augmenting
our density field with  realizations of the small scale structure
consistent with the known \iras power spectrum; the results reported
in the following sections were found to be very insensitive to
the lack of small scale resolution at large distances.

For the bias function, $b[\delta\rho]$, we consider three models:
An isotropic (I) model, where the CR source distribution is purely
isotropic, i.e. $b[\delta\rho]\equiv1$; An unbiased (UB) model, where 
the CR sources trace the galaxy distribution with
$b[\delta\rho]=1+\delta$; A biased (B) model where the CR source distribution 
is biased compared to the galaxy distribution with a simple threshold bias,
$b[\delta\rho]=1+\delta$ for $\delta > \delta_{min}$ and
zero otherwise. The value $\delta_{min}=1$ produces a 
source distribution which is concentrated near the SGP
in a manner very similar to that seen in nearby radio sources 
(Shaver \& Pierre 1989). We therefore adopt $\delta_{min}=1$ as
our canonical ``biased'' model.

We generate a specific realization of CR arrival directions
for a chosen model  by
dividing the sky into a set of angular bins; in each bin
we draw the number of sources producing 1, 2, etc. events
from  Poisson distributions with means determined by 
equation~(\ref{Sbar}). The product
$AT\dot n_0(E)$ is fixed by requiring that the mean number of CRs
produced  match the number observed by a particular
detector.  Since we are interested only in
large scale correlations of the CR sources with the galaxy
distribution, we can use relatively coarse angular binning. 
We have adopted equal
area bins of approximately $6^\circ\times6^\circ$; this
corresponds to the resolution of the \iras density field at 
100 \mpc\ and is comparable with the upper limits for
deflections caused by inter-galactic magnetic fields.
[The deflection angle for protons is (\cite{Wa})
$\theta_p\leq5\arcdeg(d/100\ {\rm Mpc})^{1/2}/(E/10^{20}{\rm eV})$
for the current upper limit on inter-galactic magnetic field $B$ with
correlation length $\lambda$,
$B\lambda^{1/2}\le10^{-9}\ {\rm G\ Mpc}^{1/2}$ (\cite{Kron}, \cite{Vallee})].
The efficiency of CR detectors is usually a function of declination. 
Relative efficiencies can easily be accounted for by multiplying the mean 
counts in (\ref{Sbar}) by the appropriate weights. 
In the case of
the Fly's eye detector, we model the efficiency as uniform above
$\delta>-10\arcdeg$ with no sensitivity at smaller declinations.

Having specified the bias model, the
only remaining parameter is the current source number density.
We have performed our Monte-Carlo realizations using two values: $\bar s_0=
10^{-2} {\rm Mpc}^{-3}$  
and $\bar s_0=10^{-4}{\rm Mpc}^{-3}$. The high number
density corresponds to the number density of bright galaxies. The low number
density was chosen based on the following argument. The energy of the
highest energy event detected by the Fly's Eye is in excess of
$2\times 10^{20}{\rm eV}$ (\cite{Fly}). 
Since the distance traveled by such a particle
can not exceed $50\ {\rm Mpc}$ (\cite{Felix}), 
its detection implies that at least one CR source
exists in the Fly's Eye field of view out to $50\ {\rm Mpc}$, implying
a source density $\bar s_0>10^{-5}\ {\rm Mpc}^{-3}$. 

A more reliable lower limit to
$\bar s_0$ may be given by considering the 
probability of observing repeating sources. As the
source number density decreases, individual sources become brighter and,
for a given number of detected events, the probability that one source
would contribute more than one event increases. 
The analysis of \S 2.1 gives an 
analytic expression for the probability to observe a repeating source
in a given angular bin. This is shown Fig. \ref{fig2} 
which gives the probability that
repeating sources would be observed (above an energy $E$) 
as a function of CR source density, $\bar s_0$,
for the Fly's Eye exposure and the isotropic model. 
From Fig. \ref{fig2}, an absence of repeaters in the current Fly's Eye data
above $5\times10^{19}{\rm eV}$ at the 90\% confidence limit  
would imply a lower limit of $\bar s_0>10^{-5}\ {\rm Mpc}^{-3}$,
similar to the lower limit inferred from the highest energy events.
Clearly, the number density would be strongly constrained by considering
repeating sources once the exposure is increased by a factor of $\sim10$.

\subsection{Results}

Figures \ref{fig3} and \ref{fig4}
present maps of the fluctuations in the mean CR intensity, 
$\Delta(E,\hat\Omega)\equiv 4\pi\bar N_o (E,\hat{\Omega})
/\int{\rm d}\hat\Omega\,\bar N_o (E,\hat{\Omega})-1$, for $E=6,\, 
4\times10^{19}{\rm eV}$ and the unbiased model.
Also shown are the SGP and the Fly's Eye field of view
for CRs with zenith angle $\theta<45\arcdeg$.\footnote[1]
{The choice of $\theta<45\arcdeg$ corresponds
to the cut made by the CR experiments: The accuracy with which the CR energy
and arrival direction is determined drops for larger zenith angle.} 
The map clearly reflects the inhomogeneity of the large-scale 
galaxy distribution- the over dense CR regions lie in the directions of 
the ``Great Attractor'' [composed of the Hydra-Centaurus 
($l=300$--$360^\circ$, $b=0$--$+45^\circ$)
and Pavo-Indus 
($l=320$--$360^\circ$, $b=-45$--$0^\circ$) superclusters]
and the Perseus-Pisces supercluster 
($l=120$--$160^\circ$, $b=-30$--$+30^\circ$).
The main structures of the CR 
arrival distribution, observable by northern hemisphere detectors such as
AGASA and the Fly's Eye, are the overdensity in the Perseus-Pisces direction 
and the underdensity along the Galactic plane. 

A crude estimate of the number
of CR events required in order to detect these structures may be obtained 
as follows. For a given detector exposure the average number of events 
expected in the under-dense, $\Delta\lesssim-0.2$, region is $20\%$ smaller
in the UB model compared to the I model. If the number of expected events in 
both models is Poisson distributed, as would be the case for
many faint sources ($\bar s_0 \rightarrow\infty$), 
then the probability that the I model would be
ruled out at a $3\sigma$ level, assuming that the source distribution 
is described by the UB model, requires that the number of detected
CRs, $m$,  satisfy (for a $1-\sigma$ upward fluctuation) 
$0.8m+(0.8m)^{1/2}\lesssim m-3m^{1/2}$, i.e.
$m\gtrsim300$. Since the $\Delta\lesssim-0.2$ region occupies $\sim1/2$
the Fly's Eye field of view, this corresponds to a total of $\sim600$
events. This number requires an exposure $\sim30$ times higher than the
current Fly's Eye exposure, for which $\sim 20$ events above 
$4\times10^{19}{\rm eV}$ were detected. A similar estimate is obtained 
considering the overdense region in the direction of the Perseus-Pisces
supercluster.

In order to obtain a more accurate estimate of the 
exposure required to discriminated between
the various models (I, UB, B) for the distribution of CR sources,
we have considered the distribution of a statistic similar
in spirit to $\chi^2$,
\begin{equation}
X(E)=\sum_l {[n_l(E)- n_{l,I}(E)]^2\over n_{l,I}(E)} \ .
\label{X}
\end{equation}
Here $n_l$ is the number of events detected in angular bin $l$ and 
$n_{l,I}$ is the average number expected for the I model. For a 
total of $\sim100$ observed CRs there would be, on average, 1 event
per $15\arcdeg\times15\arcdeg$ angular bin, limiting the angular resolution
with which the overdensity map can be reconstructed to $\gtrsim20\arcdeg$
(note, that the over/under densities predicted by the UB model are not
large). Thus,
for the calculation of $X$ we have grouped the $6\arcdeg\times6\arcdeg$
bins into $24\arcdeg\times24\arcdeg$ bins. 
In addition, we have only used
the 3 bins for which the UB model predicts the highest overdensity, and
the 3 for which the model predicts the lowest underdensity. These bins
cover $\sim20\%$ of the Fly's Eye field of view, for which
$|\Delta(E=4\times10^{19}{\rm eV},\hat\Omega)|\gtrsim0.4$ in the UB model.
It is clear from Fig. \ref{fig4} and the arguments of the previous paragraph,
that for the UB model the distribution of CRs in the remaining $\sim80\%$
of the field of view, for which $|\Delta|\lesssim0.2$, could
not be discriminated from that expected in the I model for a total of
only $\sim100$ CR events. 

The $X(E)$ distributions for the various models
are shown in Fig. \ref{fig5} 
for $E=4,\ 6,\ 8\times10^{19}{\rm eV}$ and a detector
with the Fly's Eye field of view and exposure for which $290$ CRs above 
$4\times10^{19}{\rm eV}$ are detected,
corresponding to $\sim10$ times the current Fly's Eye exposure. The detector
efficiency was assumed to be uniform within the field of view (corresponding
to declination $>-10\arcdeg$). Clearly, there is a significant probability 
that this exposure would allow to discriminate between the models:
The probability
of ruling out the isotropic model at the $3-\sigma$ level, assuming that the
CR sources are distributed as in the unbiased model, is $\gtrsim 15$\%;
The probability for all other combinations (of assumed/ruled out model)
is $\gtrsim 65$\%.
It is unlikely that the $X(E)$ statistic could discriminate 
between the unbiased and isotropic models for an exposure significantly
lower than $10$ times the current exposure of the Fly's Eye detector.
The biased model, however, could be discriminated from the isotropic model
with only $\sim3$ times the current Fly's Eye exposure.

\cite{Stanev} have recently noted a correlation between
the arrival directions of $E>4\times10^{19}{\rm eV}$ CR events detected
by the Haverah Park experiment and the Supergalactic plane.
Using our Monte-Carlo simulations we have calculated, following \cite{Stanev},
the probability distribution
of the absolute value and rms of the Supergalactic latitude,
$|b^{SG}|$ and $b^{SG}_{rms}$, and Galactic latitude,
$|b^{G}|$ and $b^{G}_{rms}$, for each of our bias models and 
for the two values of the source density, $\bar s_0$.
The declination efficiency was derived from the declination
distribution of the published Haverah Park events\cite{HP80} with
$E> 1\times 10^{19}{\rm eV}$ and zenith angles less than $45\arcdeg$.
The average number of events at each energy was chosen to correspond to
that given in \cite{Stanev}: 27 and 12 events above 
$4$ and  $6\times10^{19}{\rm eV}$ respectively. 
Table~1 presents the averages of $|b^{SG}|$, $b^{SG}_{rms}$, 
$|b^{G}|$, and $b^{G}_{rms}$ obtained from the Monte-Carlo simulations
of the various models, and the probability that a value smaller than 
the experimental Haverah Park value is obtained for each model.

In agreement with \cite{Stanev}, we find that the probability to obtain
the Haverah Park results assuming an isotropic CR source distribution is very
low. [The slight differences between our numbers for the isotropic (I)
model and those in \cite{Stanev} are due to the fact that we used a 
declination efficiency map and not the (unpublished) 
declinations of the observed events
used in \cite{Stanev}].  However, we find that the probability to obtain
the Haverah Park results is not
significantly higher for models where the CR source distribution traces 
the LSS; thus, contrary to the claim by \cite{Stanev}, we find
that the concentration of the Haverah Park events towards the SGP
is {\it not} strong support for the CR sources tracing the known LSS. 
It is important to note that for the biased
model, where CR sources are restricted to high density regions in a way
similar to radio galaxies, the probability to obtain the Haverah Park 
results is smaller than for the unbiased one. This reflects the fact that
the superclusters, while concentrated towards the SGP, 
have offsets above and below the SGP which cause the inferred CR distribution
to be less flattened than seen in the Haverah Park data. 

As seen in Fig. \ref{fig1}, there is a significant contribution to the flux
at $4\times10^{19}{\rm eV}$ from sources at distances $150-300\ {\rm Mpc}$,
where the resolution of the galaxy density field inferred from the \iras
catalogue is fairly poor. The concentration of the Haverah park events
toward the SGP might therefore be indicative of LSS at low 
Supergalactic latitude which is not seen in the \iras galaxy distribution.
In an effort to quantify this, we inserted an artificial supercluster
at $r=150$\mpc\ in the SGP with mass and size of the 
Shapley supercluster (Raychaudhury {\it et al.} 1991). 
This is most likely a worst case scenario since the actual Shapley
cluster is seen (albeit outside the Haverah Park field of view) at a similar
distance in the \iras catalog and it is unlikely that any missed
structure would lie precisely in the SGP. Even with such an extreme
structure in the SGP, all of the models were unable to reproduce
the concentration of the Haverah Park data at $\gtrsim 90$\% confidence level.
We conclude that the Haverah Park events either arise from sources
concentrated in the SGP which are not probed by the known LSS or that
there is some unknown systematic error in the positional accuracy of
the CR events. 

\section{Discussion}

We have shown that, if the distribution of CR sources trace the large scale 
structure, large exposure CR detectors should clearly reveal
anisotropy in the arrival direction distribution of CRs above
$4\times10^{19}{\rm eV}$.
The exposure required for a northern hemisphere detector to discriminate
between isotropic CR source distribution and an unbiased distribution that 
traces the LSS is approximately $10$ times the current Fly's Eye exposure.
If the CR source distribution is strongly biased, the required exposure
is $\sim3$ times the current. Increasing the exposure by a factor of $10$
with existing experiments (AGASA and Fly's Eye), would require
$\sim30$ years of observation (taking into 
account the fact that the AGASA experiment triggering was recently improved). 
The required observation time would be reduced if
new, larger, CR experiments become operative: $\sim10$ observation-years
would be required with the new High Resolution Fly's Eye experiment 
(\cite{hires}), which is planned to become operative in two years; 
Less than $1$ year of observation would be required
if the proposed $\sim5000\ {\rm km}^2$ 
detectors of the Auger project are built (\cite{huge1,huge2}).

An experiment with full sky coverage, such as the Auger experiment,
would allow the CR event distribution
to be analyzed conveniently in terms of a spherical multipole
decomposition. In Fig. \ref{fig6}, we show the probability distribution
for dipole and quadrupole moments of the CR distribution for a
full sky coverage detector with an exposure $\sim 100$ times the current
Fly's Eye exposure, comparable to the exposure of the Auger detectors
after a few years of operation.
The dipole moment has been defined as
$D= \langle \cos \theta \rangle$
where $\theta$ is the angle between the CR position and
fixed reference direction, taken to be $(l=270\arcdeg,\ b=30\arcdeg)$, while
the quadrupole moment is defined as 
$Q= \langle \sin^2 b^G \rangle - {{1}\over{3}}$.
From the figure, it is evident that even these low order statistics
are sufficient to discriminate between the models at a high
degree of confidence. Moreover, since the various multipole moments
are independent, the confidence level can be increased by 
including higher order moments of the CR distribution.

The anisotropy signal is not sensitive to the currently unknown number 
density of CR sources (see Figs. \ref{fig5} and \ref{fig6}). 
We have shown, that a reliable lower limit to the 
source number density,
$\bar s_0$, may be obtained by considering the 
probability of observing repeating sources. As the
source number density decreases, individual sources become brighter and,
for a given number of detected events, the probability that one source
would contribute more than one event increases. 
An absence of repeaters in the current Fly's Eye data
above $5\times10^{19}{\rm eV}$, for example,
would imply $\bar s_0>10^{-5}\ {\rm Mpc}^{-3}$ with 90\% confidence limit
(see Fig. \ref{fig2}). 
The number density would be strongly constrained by considering
repeating sources once the exposure is increased by a factor of $\sim10$.

\cite{Stanev} have recently noted that
the arrival directions of $E>4\times10^{19}{\rm eV}$ CR events detected
by the Haverah Park experiment show a concentration in the
direction of the SGP. In agreement with \cite{Stanev}, we find that the 
probability to obtain
the Haverah Park results assuming an isotropic CR source distribution is very
low. However, we find that this probability is not
significantly higher for models where the CR source distribution traces 
the LSS; thus, contrary to the claim by \cite{Stanev}, we find
that the concentration of the Haverah Park events towards the SGP
is {\it not} in strong support for the CR sources tracing the known LSS.

\begin{figure}
\plotone{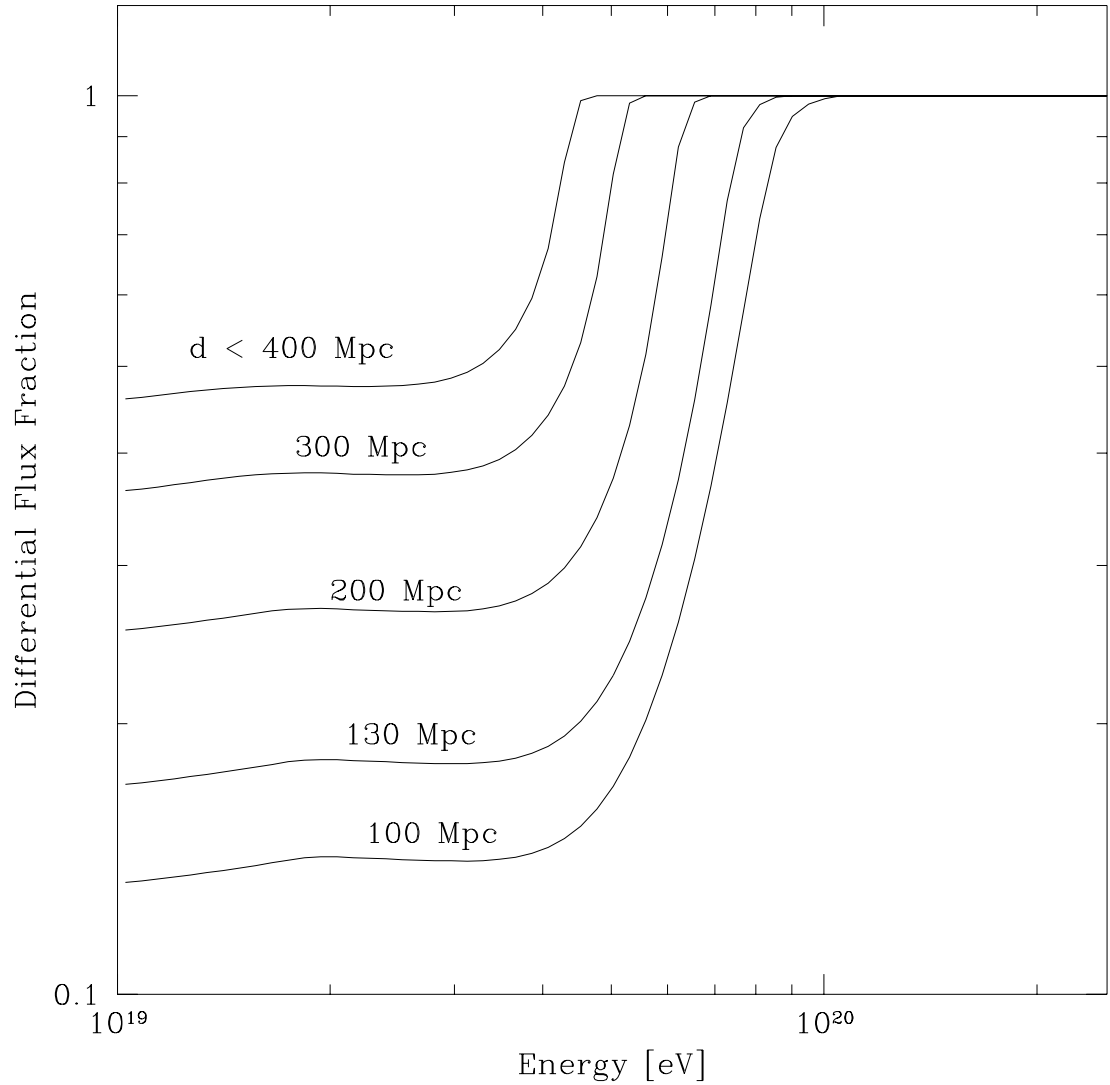}
\caption{
The fraction of the differential cosmic ray flux
contributed by a homogeneous distribution of sources with
distances $<d$ for $d=100,\ 130,\ 200,\ 300,$ and $400$ Mpc.
The sharp rise in the curves with energy arises from rapid
energy loss associated with pion production.
}
\label{fig1}
\end{figure}

\begin{figure}
\plotone{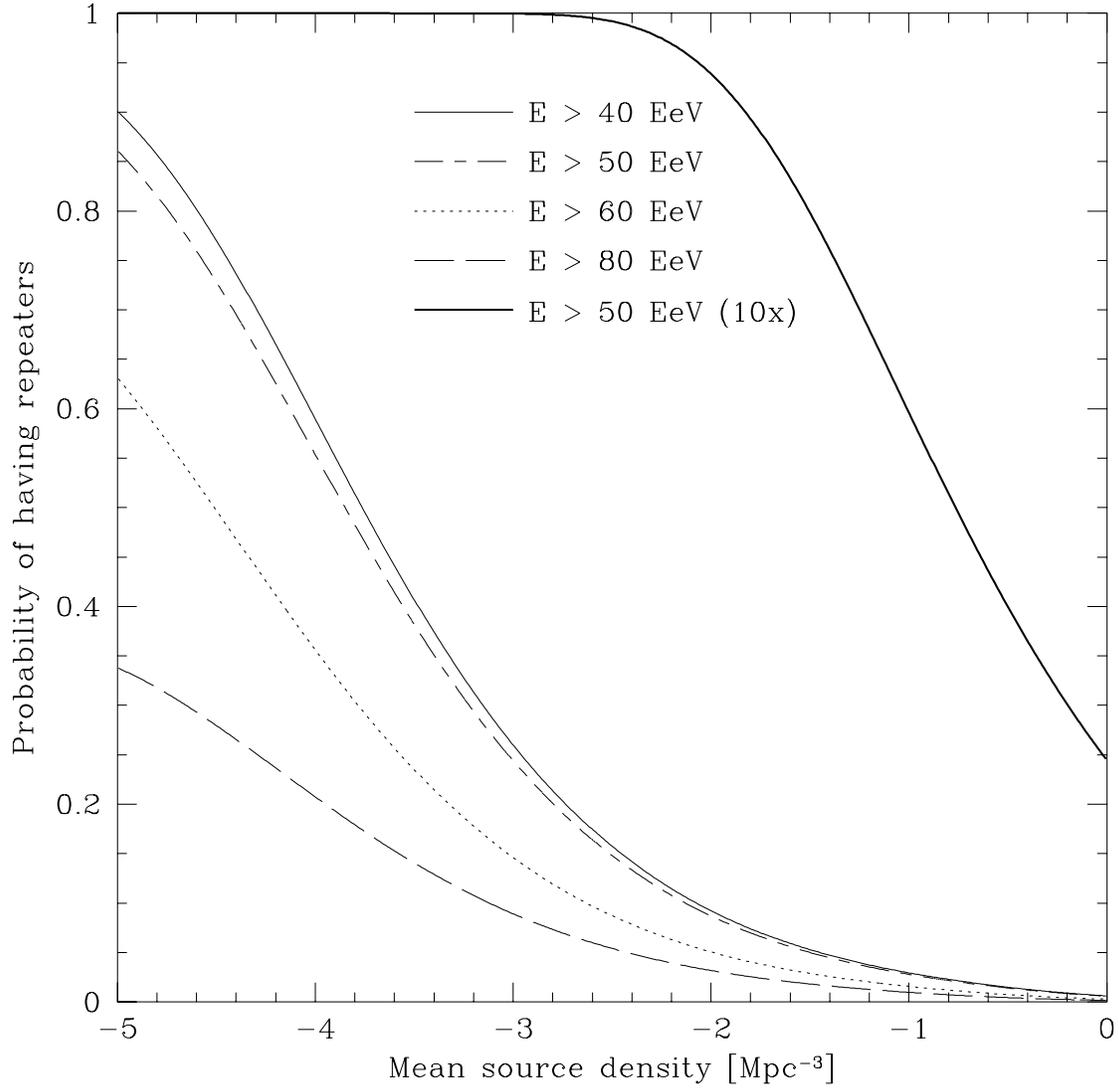}
\caption{
The probability of having sources producing more
than one CR events (repeaters) as a function of mean CR source
density for the Fly's Eye exposure. The light curves denote the probabilities
associated with cumulative energy thresholds of
40 (solid), 50 (long-short dashed), 
60 (dotted), and 80 (dashed) EeV. The heavy solid curve
denotes the probability with a threshold of 50 EeV but with
an exposure ten times the current Fly's Eye exposure.
}
\label{fig2}
\end{figure}

\begin{figure}
\plotone{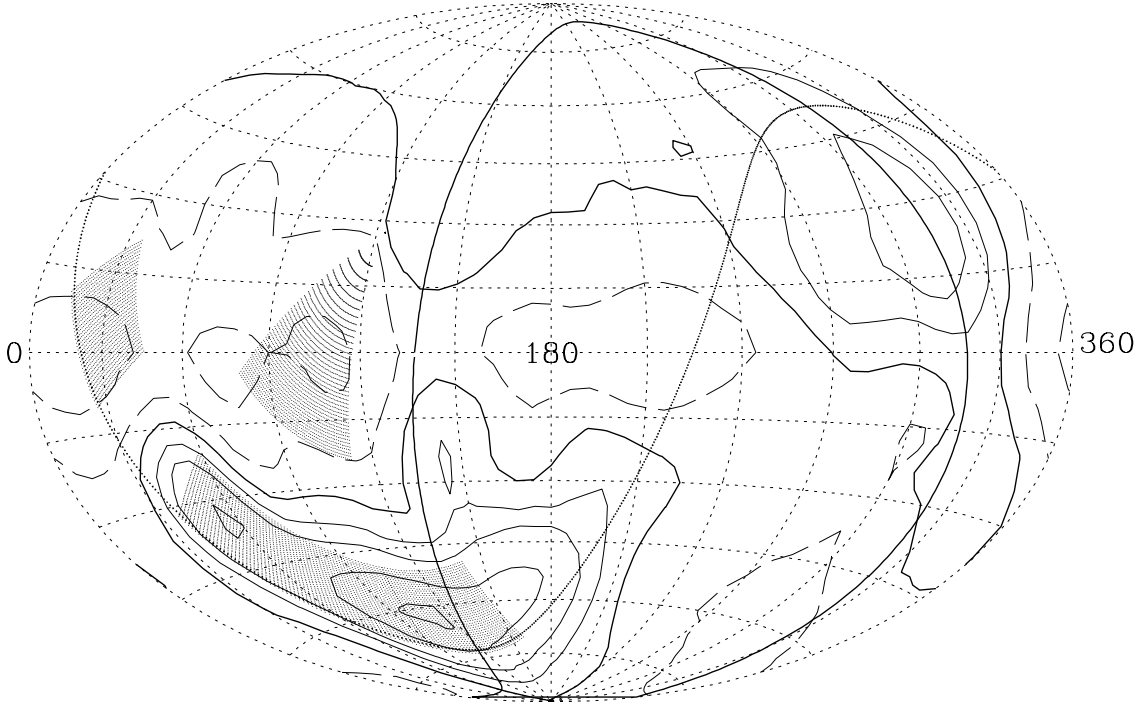}
\caption{
Aitoff projection of the fluctuations in the
average CR intensity, $\Delta(E,\hat\Omega)$, for $E=60$ Eev.
The sky coverage was taken to be uniform and the CR sources
were assumed to trace the LSS ($b[\delta\rho ] = 1 + \delta$).
The heavy curve denotes the zero contour. Solid (dashed) contours
denote positive (negative) fluctuations at intervals
$[-0.5,\ -0.25,\ 0,\ 0.25,\ 0.50,\ 1.0,\ 1.5]$. The Supergalactic
plane is denoted by the heavy solid curve roughly perpendicular
to the Galactic plane. The dotted curve denotes the 
Fly's Eye coverage of declination $>-10^\circ$. The shaded regions show
the high and low density regions used in the $X(E)$ statistic
(equation 5).
}
\label{fig3}
\end{figure}

\begin{figure}
\plotone{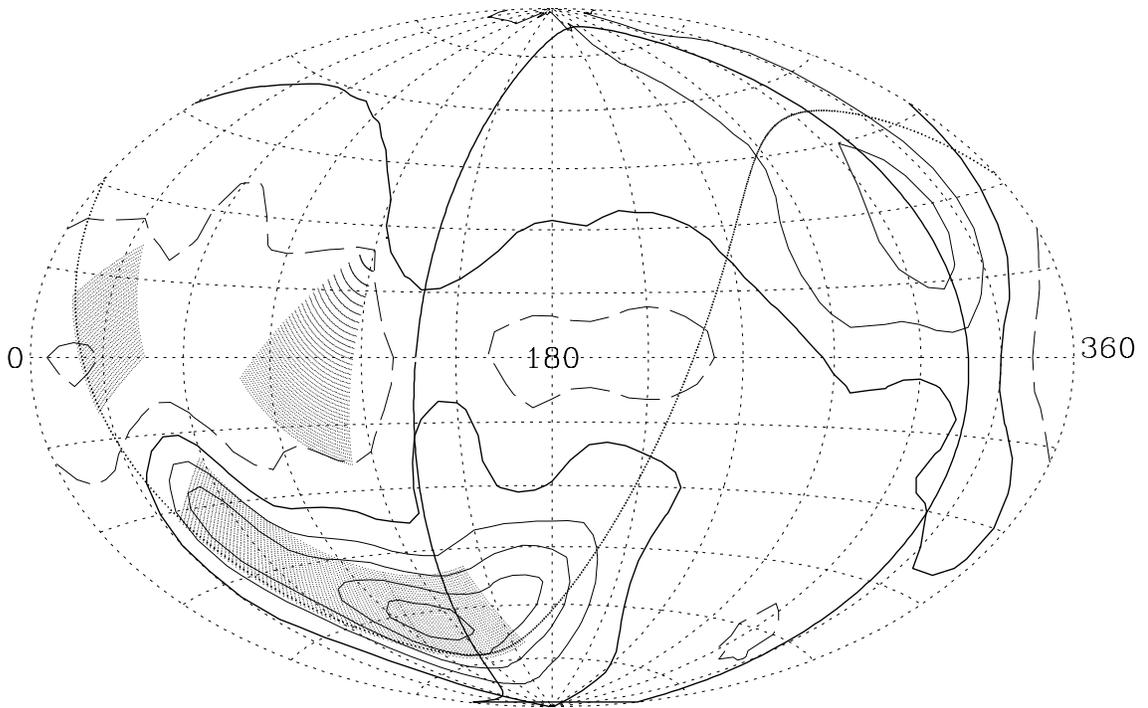}
\caption{
Same as Fig. 3, for $E=40$ EeV. 
The heavy curve denotes the zero contour. Solid (dashed) contours
denote positive (negative) fluctuations at intervals
$[-0.4,\ -0.2,\ 0,\ 0.2,\ 0.4,\ 0.6,\ 0.8]$.
}
\label{fig4}
\end{figure}

\begin{figure}
\plotone{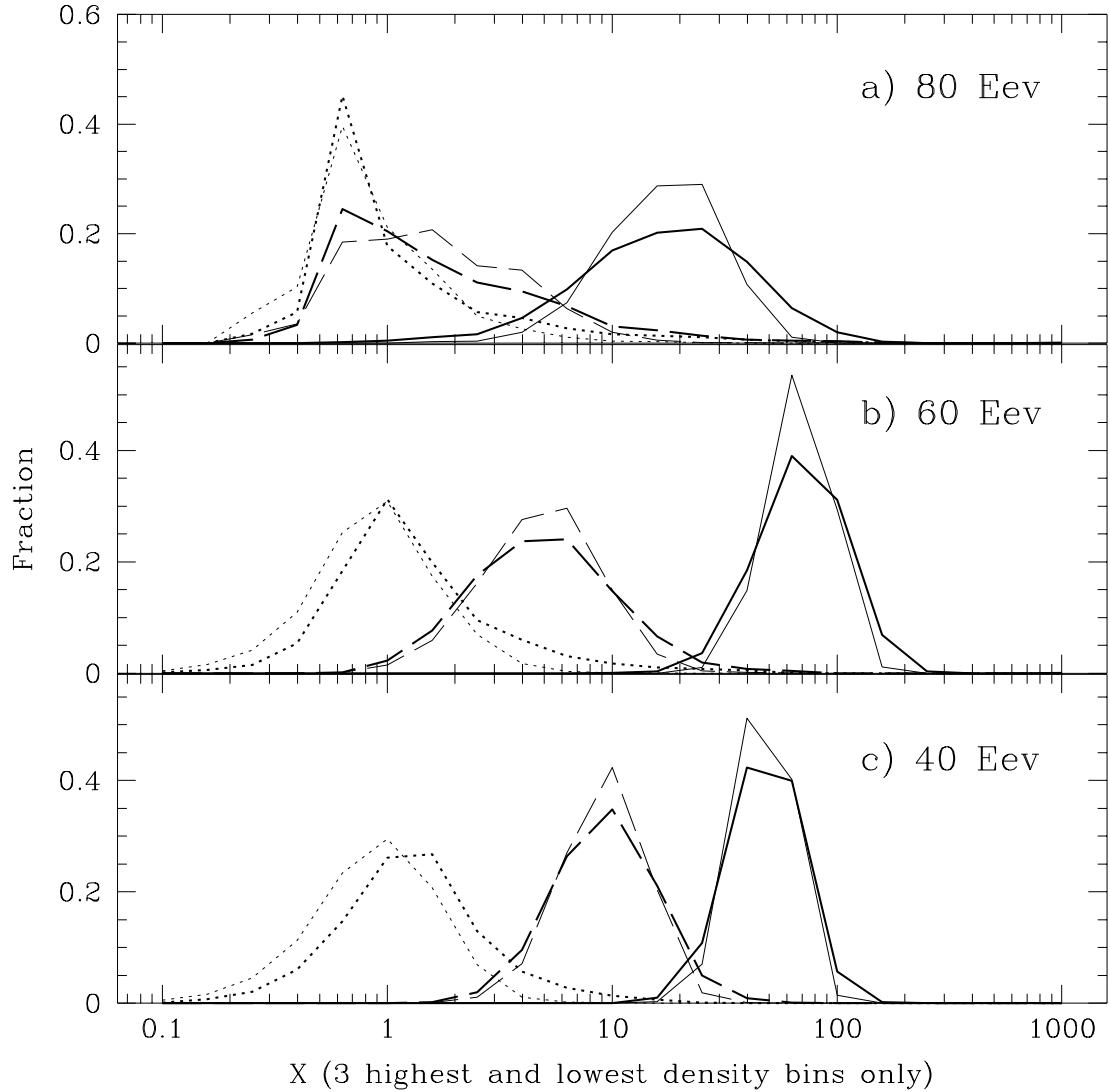}
\caption{
Probability distribution of the statistic $X(E)$
(equation 5) for cumulative energies of a) 80, b) 60, and
c) 40 Eev computed from 10,000 Monte Carlo realizations of
the Fly's Eye detector with 10 times its current exposure.
In each panel the solid curves denote the
biased model (B), while the dashed curves represent the unbiased
model (UB), and the dotted curves are the isotropic (I) model.
The light set of curves correspond to a mean CR source density of
$\bar s_0 = 10^{-2}{\rm Mpc}^{-3}$ while the heavier curves 
correspond to $\bar s_0 = 10^{-4}{\rm Mpc}^{-3}$.
}
\label{fig5}
\end{figure}

\begin{figure}
\plotone{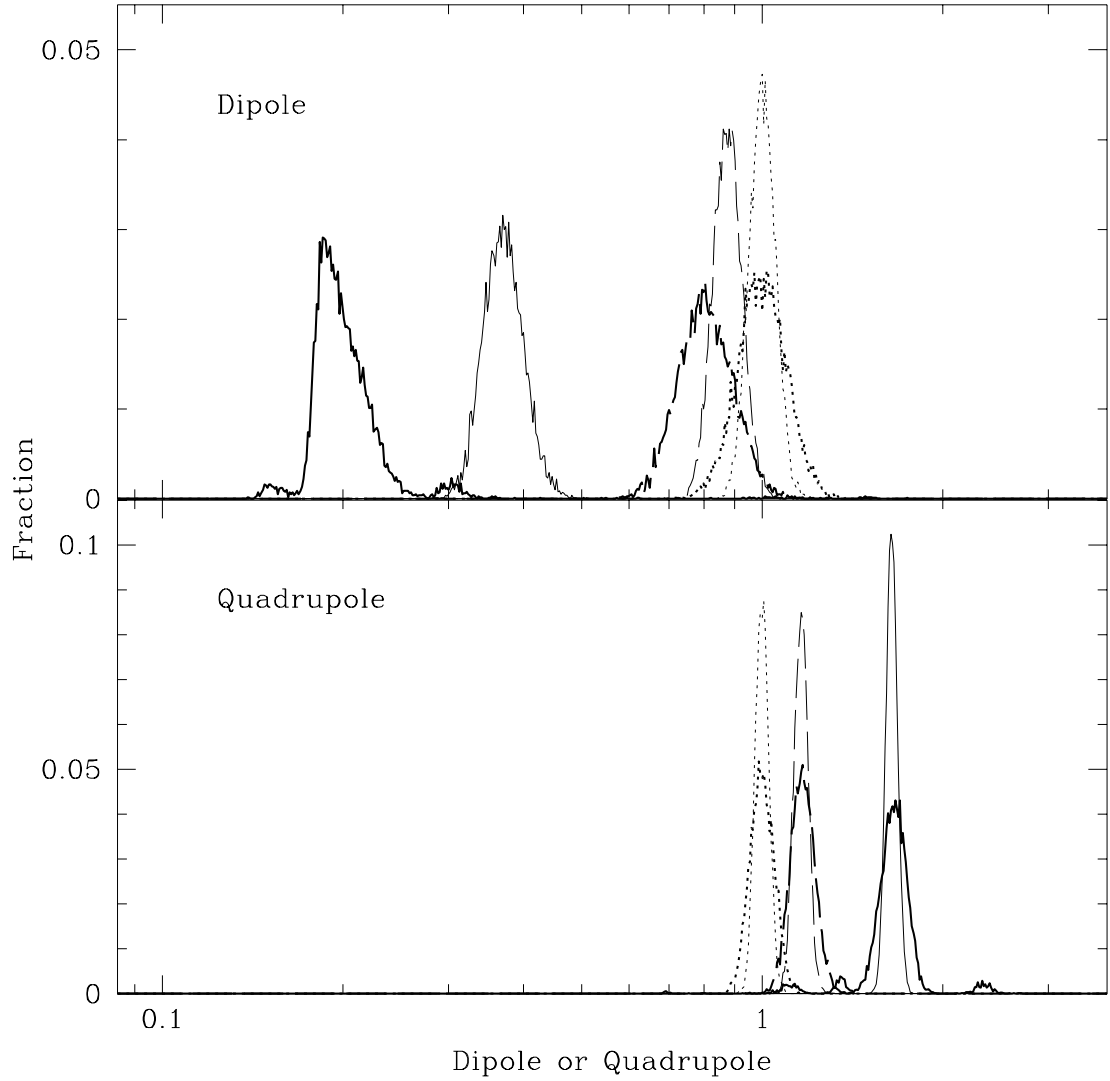}
\caption{
Probability distribution of the dipole and quadrupole
moments of the CR distribution computed from 10,000 Monte-Carlo
realizations of proposed full-sky Auger detector with an effective
exposure of 100 times the current Fly's Eye exposure and a 
cumulative energy threshold of 60 EeV.  The various
line types correspond to the models the same as in figure~4.
}
\label{fig6}
\end{figure}

\begin{figure}
\plotone{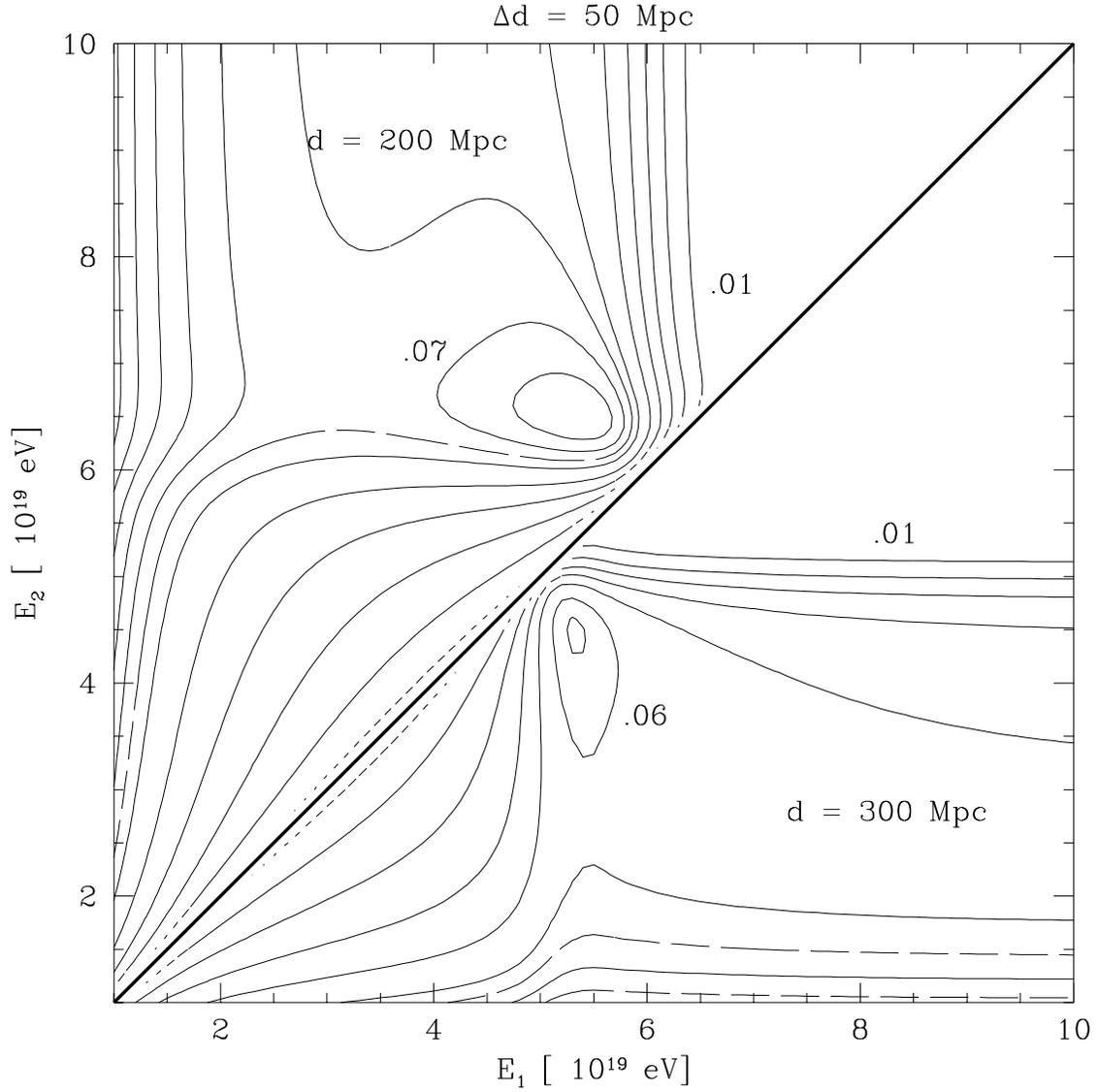}
\caption{
Signal to noise contours, $\sigma(E_1,E_2,d_1,d_2)$, for the
energy interval corresponding to $E_1$ and $E_2$ and
a distance resolution of $d_2-d_1= 50$ Mpc. The upper
panel shows the contours appropriate for probing structures
at $(d_1+d_2)/2=200$ Mpc, while the lower panel is ``tuned'' to 
$(d_1+d_2)/2=300$ Mpc. The
solid contours are spaced at $\Delta\sigma=0.01$. The dashed
contours are at $\sigma=0.065$ and $0.075$.  
}
\label{fig7}
\end{figure}

Our analysis addressed only the 2-dimensional (angular) LSS information
contained in the distribution of CR arrival directions.
However, the energy dependent distance cutoff of high energy CRs, shown
in Fig. \ref{fig1}, 
implies that the differential CR flux is dominated at different
energies by sources which lie at different distances. Therefore, analyzing
the angular distribution of CRs at different energy ranges would provide
information on the LSS at different distances, and would therefore
probe the 3-dimensional LSS. The exposure which would be required in
order to extract 3-dimensional LSS information from the CR arrival 
distribution may be estimated as follows. Let us assume that we are
interested in a LSS feature that lies within a distance range $d_1-d_2$
and occupies a solid angle $\delta\Omega$. 
Let's denote the fraction of the flux in the
energy range $E_1-E_2$, that is contributed on average (i.e. for homogeneous
source distribution) by sources within the distance range $d_1-d_2$,
by $f(E_1,E_2,d_1,d_2)$. Our ``signal'' in the energy range $E_1-E_2$, 
i.e. the number of CRs in this energy range that are produce by sources
that lie within the LSS feature of interest, is then
given by $\sim f(E_1,E_2,d_1,d_2)
N(E_1,E_2)\delta\Omega/{\rm d}\Omega$, where $N(E_1,E_2)$
is the total number of events in the $E_1-E_2$ range and ${\rm d}\Omega$ is 
the experimental field of view. 
The "noise" from sources outside $d_1-d_2$ is 
$\sim[(1-f)N\delta\Omega/{\rm d}\Omega]^{1/2}$, so that the signal to noise is
$\sigma(E_1,E_2,d_1,d_2)\simeq f[N\delta\Omega/{\rm d}\Omega
(1-f)]^{1/2}$. Figure \ref{fig7} shows a contour map of the signal to 
noise $\sigma(E_1,E_2)$ for a structure with radial extent of
$50{\rm Mpc}$ and angular extent of $10\arcdeg\times10\arcdeg$ lying at a 
distance of $d=(d_1+d_2)/2=200,\,300{\rm Mpc}$. For this figure, the exposure 
was chosen to be the current Fly's Eye exposure. The map shows that, as 
indicated by Fig. \ref{fig1}, the best signal to noise is obtained by choosing
the energy range $5.5-6.5\times10^{19}{\rm eV}$ for $d=200{\rm Mpc}$,
and $4.5-5.5\times10^{19}{\rm eV}$ for $d=300{\rm Mpc}$. Unfortunately, the 
exposure required to obtain a signal to noise of order $1$ is very
large, $\sim300$ times the current Fly's Eye exposure, which correspond
to $\gtrsim10$ observation years of the proposed Auger detectors. Thus, 
although it is possible in principle to probe the 3-dimensional LSS 
using UCHERs at different energy channels, it is not clear if it would be
possible to do so in practice, even with the largest detectors planned today.

\acknowledgements
We thank Jordi Miralda-Escud\'e and John Bahcall  
for valuable suggestions and comments, A. A. Watson for information on
the Haverah Park experiment, and P. Sommers for information on the Fly's
Eye experiment. EW thanks the Institute for Theoretical Physics
(UC Santa-Barbara) for its hospitality.
This research was partially supported by a W. M. Keck Foundation grant 
and NSF grants PHY 95-13835, PHY94-07194.

\clearpage

\begin{deluxetable}{lccccc}
\tablecaption{Monte Carlo Results}
\scriptsize
\tablehead{
\colhead{Model} & 
\colhead{$>E$ Eev} &
\colhead{$b_{RMS}^{G}$} &
\colhead{$b_{RMS}^{SG}$}  &
\colhead{$\langle |b^G|\rangle$} &
\colhead{$\langle|b^{SG}|\rangle$ } }
\startdata 
Data & 40 & 46.5 \ \ --- & 23.3 \ \ --- & 39.5 \ \ --- &  18.6 \ \ ---    \nl
I 
($\bar s_0 = 10^{-2}$) 
&40 & 37.3 \ \ 0.980 & 33.8 \ \ 0.007 & 30.1 \ \ 0.984 & 27.3 \ \  0.010\nl
I  
($\bar s_0 = 10^{-4}$) 
&40 & 37.2 \ \ 0.971 & 33.8 \ \ 0.009 & 30.0 \ \ 0.974 & 27.4 \ \ 0.017\nl
UB  
($\bar s_0 = 10^{-2}$)  
&40 & 40.4 \ \ 0.926 & 31.5 \ \ 0.023 & 33.9 \ \ 0.907 & 25.2\ \ 0.032\nl
UB 
($\bar s_0 = 10^{-4}$) 
&40 & 40.4 \ \ 0.895 & 31.4 \ \ 0.032 & 34.0 \ \  0.875 & 25.2 \ \ 0.050\nl
B  
($\bar s_0 = 10^{-2}$) 
&40 & 38.6 \ \ 0.961 & 32.1 \ \ 0.017 & 31.3 \ \ 0.970 & 25.5 \ \ 0.027\nl
B 
($\bar s_0 = 10^{-4}$) 
&40 & 38.0 \ \ 0.947 & 32.2 \ \ 0.021 & 30.9 \ \ 0.951 & 25.7 \ \ 0.035\nl
   &  &   &             &    &   \nl
Data & 60 & {52.7}  \ \ --- & 23.9  \ \ ---  & 45.9  \ \ --- & 18.2 \ \ --- \nl
I 
($\bar s_0 = 10^{-2}$) 
&60 & 38.0 \ \ 0.977 & 34.5 \ \ 0.050 & 30.1 \ \ 0.987 & 27.4 \ \ 0.053\nl
I 
($\bar s_0 = 10^{-4}$) 
&60 & 38.1 \ \ 0.955 & 34.5 \ \ 0.083 & 30.2 \ \ 0.972 & 27.4 \ \ 0.085\nl
UB 
($\bar s_0 = 10^{-2}$) 
&60 & 42.8 \ \ 0.922 & 29.8  \ \ 0.170 & 35.7  \ \ 0.938 & 23.3 \ \ 0.169\nl
UB 
($\bar s_0 = 10^{-4}$) 
&60 & 43.0 \ \ 0.876 & 29.6 \ \ 0.227 & 35.8 \ \ 0.895 & 23.1 \ \ 0.228\nl
B 
($\bar s_0 = 10^{-2}$) 
&60 & 47.5 \ \ 0.729 & 16.7 \ \ 0.957 & 39.1 \ \ 0.806 & 13.2 \ \ 0.944\nl
B 
($\bar s_0 = 10^{-4}$) 
&60 & 38.5 \ \ 0.886 & 17.9  \ \ 0.863 & 31.5 \ \ 0.920 & 14.1 \ \ 0.822\nl
\enddata 
\tablenotetext{}{Numbers to the right reflect the percentage of
10,000 Monte-Carlo runs which had values as low the data.}

\end{deluxetable}

\clearpage

\end{document}